\documentclass[useAMS,usenatbib]{mn2e}
\usepackage{graphics,rotating,multirow,bm,color,hyperref}
\hypersetup{
    colorlinks=true,
    linkcolor=black,
    citecolor=black,
}

\title[Halos and Voids in $f(R)$ gravity]
  {Halos and Voids in $f(R)$ gravity}
\author[Baojiu~Li, Gong-Bo~Zhao and Kazuya~Koyama]
  {Baojiu~Li$^{1,2,3,4,}$\thanks{E-mail: baojiu.li@durham.ac.uk}, Gong-Bo~Zhao$^{5,}$\thanks{E-mail: gong-bo.zhao@port.ac.uk} and Kazuya~Koyama$^{5,}$\thanks{E-mail: kazuya.koyama@port.ac.uk}\\
  $^1$Institute for Computational Cosmology, Department of Physics, University of Durham, South Road, Durham DH1 3LE, UK\\
  $^2$DAMTP, Centre for Mathematical Sciences, University of Cambridge, Wilberforce Road, Cambridge CB3 0WA, UK\\
  $^3$Kavli Institute for Cosmology Cambridge, Madingley Road, Cambridge CB3 0HA, UK\\
  $^4$Institute of Astronomy, Madingley Road, Cambridge CB3 0HA, UK\\
  $^5$Institute of Cosmology and Gravitation, University of Portsmouth, Dennis Sciama Building, Portsmouth PO1 3FX, UK}

\pagerange{\pageref{firstpage}--\pageref{lastpage}} \pubyear{2011}

\def\LaTeX{L\kern-.36em\raise.3ex\hbox{a}\kern-.15em
    T\kern-.1667em\lower.7ex\hbox{E}\kern-.125emX}

\def\ie{{\frenchspacing\it i.e.}}

\newcommand{\tc}[1]{\textcolor{blue}{#1}}

\newcommand{\m}{{\rm M}}

\newcommand{\fR}{f_{R}}
\newcommand{\dfR}{{\delta}f_{R}}

\newcommand{\rd}{{\rm d}}
\newcommand{\dr}{\delta{R}}
\newcommand{\remove}[1]{}

\def\ie{{\frenchspacing\it i.e.}}

\def\be{\begin{equation}}
\def\ee{\end{equation}}
\def\ba{\begin{eqnarray}}
\def\ea{\end{eqnarray}}

\begin{document}

\label{firstpage}

\maketitle

\begin{abstract}
In this paper, we study the distribution of dark matter halos and voids using high resolution simulations in $f(R)$ gravity models with the chameleon mechanism to screen the fifth force in dense environment. For dark matter halos, we show that the semi-analytic thin shell condition provides a good approximation to describe the mass and environmental dependence of the screening of the fifth force in halos. Due to stronger gravity, there are far more massive halos and large voids in $f(R)$ models compared with $\Lambda$CDM models. The numbers of voids with an effective radius of 15$h^{-1}$Mpc are twice and four times as many as those in $\Lambda$CDM for $f(R)$ models with $|f_{R0}|=10^{-5}$ and $10^{-4}$ respectively. This provides a new means to test the models using the upcoming observational data. We also find that halos inside voids are all unscreened in our simulations, which are ideal objects for the gravity test.

\end{abstract}


\section{Introduction}

\label{sect:intro}

The biggest problem in cosmology is to explain the observed accelerated expansion of the universe. Within the framework of General Relativity (GR), the acceleration originates from dark energy \citep{cst2006}. Alternatively, a large-distance modification to GR may account for the late-time acceleration of the universe.

It has been recognised that usually once we modify Einstein gravity on large scales, there can appear a new scalar degree of freedom in gravity which mediates a fifth force. Without a mechanism to suppress this additional force, most modified gravity models are excluded by stringent constraints on deviations from GR on the solar system scale. One way to evade these constraints is to exploit a chameleon mechanism \citep{kw2004}. The new scalar degree of freedom couples to the energy density of matter. By tuning the coupling and potential for the scalar mode, it is possible to realise a situation that in dense environments such as the solar system, the scalar field has a large mass and it essentially does not mediate the fifth force. On the other hand, on cosmological scales, this scalar mode is light and modifies gravity significantly.

In models with the chameleon mechanism, there appears environmental dependence on the properties of dark matter distributions. In dense environment such as in clusters, the chameleon works efficiently and the modification of gravity is suppressed. On the other hand, in underdense regions such as in voids, the chameleon mechanism does not work and gravity is significantly modified. As is shown in our previous paper \citep{zlk2011b}, this environmental dependence is a smoking gun for the modification of gravity in models with the chameleon mechanism.

In this paper, we study the properties of dark matter halos and voids in models with the chameleon mechanism to reveal the environmental dependence of dark matter distributions. We use $f(R)$ gravity as an example and exploit recent results from high resolution simulations described in \citet{zlk2011}. In the $f(R)$ gravity, the fifth force can enhance gravity by a factor of $1/3$ but this enhancement is suppressed by the chameleon mechanism in the overdense regions. Some other numerical simulations performed for models with chameleon mechanism are those in \citet{oyaizu2008,olh2008,sloh2008,lz2009,zmlhf2010,lz2010} and \citet{lztk2011}. Although we will not study those simulations directly, we expect that the results found here will be true for them as well.

The paper is organised as follows. In Section II, we summarise $f(R)$ gravity models that we shall study in this paper. In Section III, we study the probability distribution function of the smoothed density field and show how the properties of halos and voids in $f(R)$ gravity are modified compared with the standard $\Lambda$CDM model. Section IV is devoted to the study of dark matter halos. We study how the difference between dynamical and lensing masses, which arises due to the fifth force, depends on the mass and environment of halos. We find that the semi-analytic thin shell condition that determines the efficiency of the chameleon can well describe those dependence found in simulations. In Section V, we study the underdense regions by identifying voids in our simulations. We study the properties of halos inside and near the voids. We show that the number density of large voids is significantly modified in $f(R)$ gravity models.

\section{$f(R)$ Gravity and Simulations}

\label{sect:model}

The $f(R)$ gravity, in which the Ricci scalar $R$ in the Einstein-Hilbert action is generalised to a function of $R$, was designed to explain the observed cosmic acceleration without introducing dark energy. In such theories, the structure formation is determined by the following equations,

\begin{eqnarray}
\label{eq:Poisson} \nabla^2 \Phi&=&\frac{16\pi{G}}{3}a^2\delta\rho_{\m}+\frac{a^2}{6}\dr(\fR),\\
\label{eq:sf} \nabla^2\dfR &=&-\frac{a^2}{3}[\dr(\fR)+8\pi{G}\delta\rho_{\m}],
\end{eqnarray}
where $\Phi$ denotes the gravitational potential, $\fR\equiv\frac{\rd f(R)}{\rd R} $ is the \emph{scalaron}, the extra scalar degree of freedom, $\dr=R-\bar{R},\delta\rho_{\m}=\rho_{\m}-\bar{\rho_{\m}}$, and the quantities with overbar take the background values. In GR, the gravitational potential is solely determined by the distribution of matter, say,  $\nabla^2 \Phi=4\pi{G}a^2\delta\rho_{\m}$. This is a linear Poisson equation, which is much easier to solve. In $f(R)$, however, the scalar field complicates the Poisson equation, making the effective Newton's constant vary with the local density: in underdense regions, the $\dr(\fR)$ term in Eq.~(\ref{eq:Poisson}) vanishes thus two equations decouple, making the effective Newton's constant enhanced by $1/3$. On the other hand, in the dense region, $\dfR$ in Eq.~(\ref{eq:sf}) is negligible, so $\dr(\fR)=-8\pi{G}\delta\rho_{\m}$, which means that GR is locally restored. This is the chameleon mechanism making $f(R)$ evade the stringent solar system tests, thus is important for the cosmological viability of the $f(R)$ gravity.

One could rewrite Eq.~(\ref{eq:Poisson}) as, \be \nabla^2\Phi = 4\pi{G}a^2\delta\rho_{\rm eff},\ee where the effect of the scalar field is absorbed into the definition of the effective energy density $\delta\rho_{\rm eff}$. Then the dynamical mass $M_D(r)$ of a halo is defined as the mass
contained within a radius $r$, inferred from the gravitational
potential felt by a test particle at $r$. It is given by $M_{D}\equiv\int{a^2}\delta\rho_{\rm eff}dV$, in which the integral
is over the extension of the body.
On the other hand, the lensing mass is the true mass of the halo, \ie, $M_{L}\equiv\int{a^2}\delta\rho_{\rm M}dV$.

Comparing the lensing mass with the dynamical mass of the same halo can in principal be a easy way to test GR. This is because the lensing mass and the dynamical mass are identical in GR, but
quite different in MG scenarios. To quantify
the difference, we calculate the relative difference $\Delta_M$
between $M_{L}$ and $M_{D}$ for each halo, $ \Delta_M\equiv
M_{D}/M_{L}-1$. Note that in $f(R)$, $ \Delta_M\le1/3$ \citep{zlk2011b}.

The presence of the chameleon effect indicates that Eqs (\ref{eq:Poisson}) and (\ref{eq:sf}) are highly nonlinear, so that the system cannot be solved without using N-body simulations. In this work, we we shall use the high-resolution $N$-body simulation catalogue \citep{zlk2011}
for a $f(R)$ gravity model, $f(R)=\alpha{R}/(\beta{R}+\gamma)$ \citep{hs2007}
where $\alpha=-m^2{c_1},\beta={c_2},\gamma=-m^2,m^2=H_0^2\Omega_{\rm M}$ and $c_1,c_2$ are free parameters. The
expansion rate of the universe in this $f(R)$ model is determined
by $c_1/c_2$, and the structure formation depends on $|f_{R0}|$,
which is the value of $|df/dR|$ at $z=0$, and is proportional
to $c_1/c_2^2$. We tune $c_1/c_2$ to obtain the same expansion
history as that in a $\Lambda$CDM model, and choose values for
$|f_{R0}|$ so that those models cannot be ruled out by current
solar system tests. To satisfy these requirements, we set
$c_1/c_2=6\Omega_{\Lambda}/\Omega_{\rm M}$ and simulate three
models with $|f_{R0}|=10^{-4},10^{-5},10^{-6}$.
In this paper, we study the distribution of dark matter halos and voids in these simulations at $z=0$. The method to identify halos is described in \citet{lb2011,zlk2011}.

\section{Probability Distribution of Density Field}

\label{sect:dens_prob}

\begin{figure}
\includegraphics[scale=0.35]{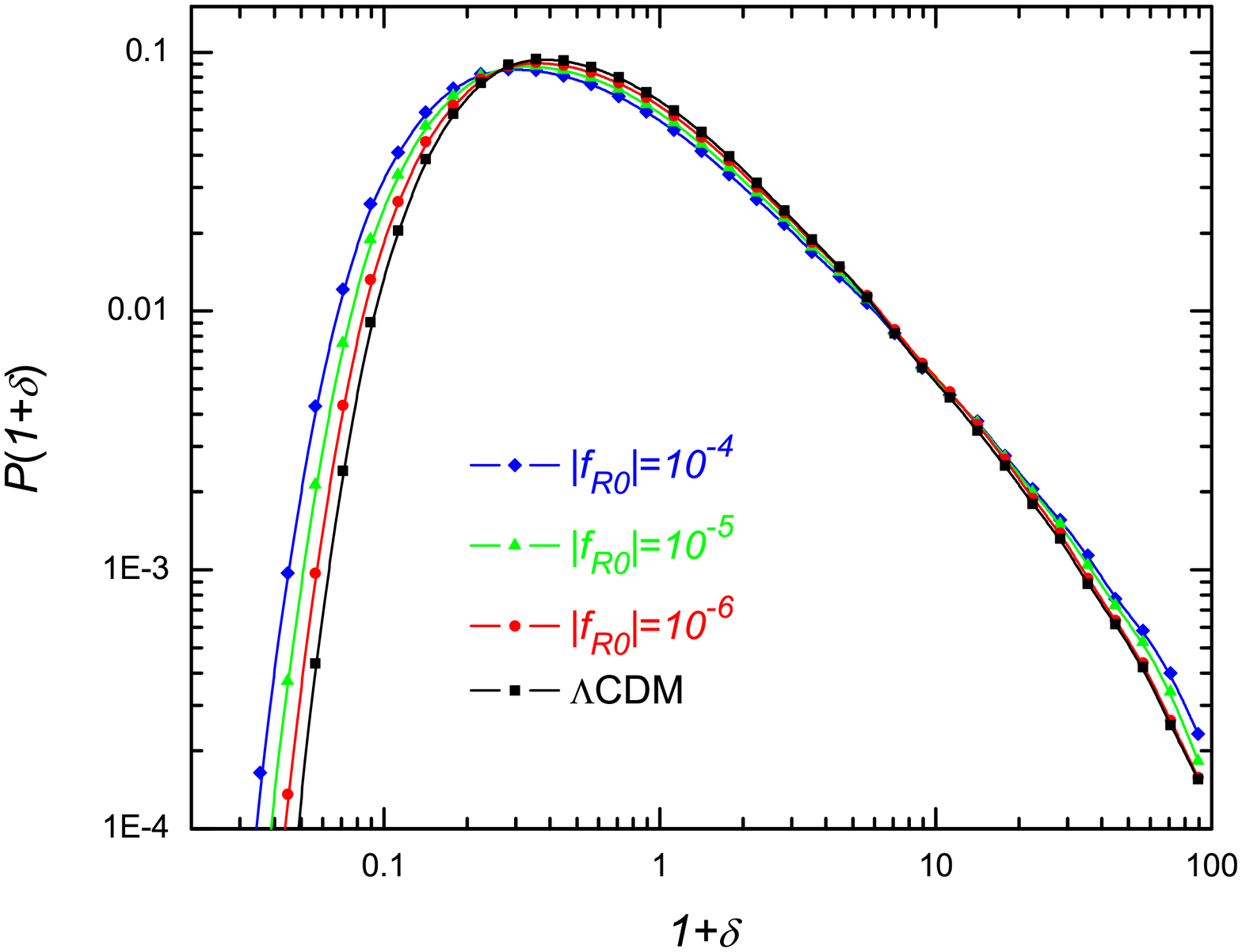}
\caption{(Colour Online) The probability distribution of matter density contrast field $\delta\equiv\rho({\bf x};R)/\bar{\rho}-1$. The $\delta$ field was filtered by a top-hat window with radius $R=2h^{-1}$Mpc. In the plot we offset $\delta$ by 1 for the ease of using the logarithmic scale and the points plotted homogeneously in $\lg(1+\delta)$. The black squares, red circles, green triangles and blue diamonds are from the $\Lambda$CDM simulation and $f(R)$ simulations with $|f_{R0}|=10^{-6}$, $10^{-5}$, $10^{-4}$ respectively. Each curve represents the averaged result over ten realisations and is normalised so that the integration of $P(1+\delta)$ is $1$.}
\label{fig:dens_prob}
\end{figure}

In the standard cold dark matter paradigm, structures grow from the small inhomogeneities in the initial matter density field due to the pull of gravity. As a result, initial overdense (underdense) regions become more and more overdense (underdense). In $f(R)$ gravity, gravity can be enhanced, so that the fifth force helps to pull more matter into overdense regions, and the underdense regions can be evacuated more efficiently.

In Fig.~\ref{fig:dens_prob} we show the probability distribution of the matter density contrast field measured from our $f(R)$ and $\Lambda$CDM simulations. This is calculated by filtering the density field by top-hat windows with as an example radius $R=2h^{-1}$Mpc centred at each cell of the simulation grid, and counting how many such windows fall into a given density band.

As shown in this figure, the fifth force can tremendously increase the chance of creating extremely low-density regions in the Universe. For example, only $\sim0.01\%$ of the space in the $\Lambda$CDM paradigm has a density of $1+\delta=0.05$, while the $f(R)$ models predict $5$ (for $|f_{R0}|=10^{-6}$), $15$ (for $|f_{R0}|=10^{-5}$) and $30$ (for $|f_{R0}|=10^{-4}$) times as much. The effect becomes smaller for increasing $\delta$, and for $1+\delta\sim0.2-0.3$ the probability becomes roughly the  same for all models.  For $0.2-0.3<1+\delta<\sim10$ the fifth force actually decreases the probability and then for windows with $1+\delta>\sim10$ the fifth force makes it more likely to be found again by making matter cluster more strongly. Similar effects have been found for other models \citep{hj2009, li2011}.

The peaks of the density distribution shifts towards low values as $|f_{R0}|$ increases, which shows that the Universe in $f(R)$ gravity may look emptier overall. Meanwhile, Fig.~\ref{fig:dens_prob} confirms that a $f(R)$ universe will more likely host very big voids and very massive dark matter halos, both of which are rarer in a $\Lambda$CDM universe. We will come back to this point later.

Fig.~\ref{fig:dens_prob} also clearly shows the effect of the chameleon mechanism, which is known to work better for smaller $|f_{R0}|$ and for high density fields \citep{zlk2011}. For $|f_{R0}|=10^{-6}$, the deviation from LCDM is suppressed for high density fields while there is still a sizable deviation in the probability distribution for under-density fields. This shows that the modification of gravity is more prominent for voids for small $|f_{R0}|$. This fact is important when we perform observational tests of modified gravity models with a realistic value of $|f_{R0}|$ compatible with the solar system constraints.

\section{Overdense Regions}

\label{sect:ovdens}

In general, dark matter halos reside in high-density regions, which form their local environment. It is well known that the fifth force in $f(R)$ gravity sensitively depends on the environment. Thus from a theoretical point of view\tc{,} it is very important to understand how the environment changes the properties of the fifth force.

As mentioned above, $f(R)$ gravity is a subclass of the chameleon scalar field theory, with $f_R=\exp(\gamma\sqrt{\kappa}\varphi)-1$, in which $\kappa=8\pi G=M_{\rm Pl}^{-2}$ and $\varphi$ is the corresponding scalar field and $\gamma=\sqrt{2/3}$ is the constant coupling strength. The scalar field is governed by an effective potential (see, {\it e.g.}, \citet{lb2007})
\begin{equation}\label{eq:Veff}
V_{\rm eff}(\varphi) = \frac{Rf_R-f}{2\kappa\left(1+f_R\right)^2}+\frac{1}{4}\rho_m\exp\left(\gamma\sqrt{\kappa}\varphi\right).
\end{equation}
When the chameleon mechanism is at work, a spherical body will develop a thin shell, and the thickness of which is given by (\citet{kw2004,le2011}),
\begin{eqnarray}\label{eq:thin_shell}
\frac{\Delta R}{R} = \frac{\varphi_{\rm out}-\varphi_{\rm in}}{\gamma\sqrt{\kappa}\rho_{\rm in}R^2},
\end{eqnarray}
in which $R$ is the radius of the body, $\Delta R$ is the thickness of the shell, $\varphi_{\rm out}, \varphi_{\rm in}$ are the values of $\varphi$ minimising $V_{\rm eff}$ inside and outside the body, respectively. Similarly, $\rho_{\rm in}$ and $\rho_{\rm out}$ are the constant matter density inside and outside the body respectively. Note that only the matter inside the thin shell contributes to the fifth force exerting on a nearby test particle. From Eq.~(\ref{eq:thin_shell}) it is evident that the fifth force could be suppressed if shell becomes thinner, which can be realised by the following two ways:
\begin{enumerate}
\item Increasing $R$ and/or $\rho_{\rm in}$, thereby making the body (in the case here the dark matter halo) more massive.
\item Decreasing $\varphi_{\rm out}$, which involves increasing $\rho_{\rm out}$ or equivalently making the environment denser\footnote{Note that the first term on the right-hand side of Eq.~(\ref{eq:Veff}) is a runaway potential of $\varphi$, while the second term increases exponentially in $\varphi$. So $V_{\rm eff}(\varphi)$ has a global minimum, which shifts towards smaller values of $\varphi$ when $\rho_m$ increases.}.
\end{enumerate}
As a result, massive halos in dense environments are strongly screened from the fifth force, while small halos in low-density environments are less screened and may experience the full fifth force.

\begin{figure*}
\includegraphics[scale=0.45]{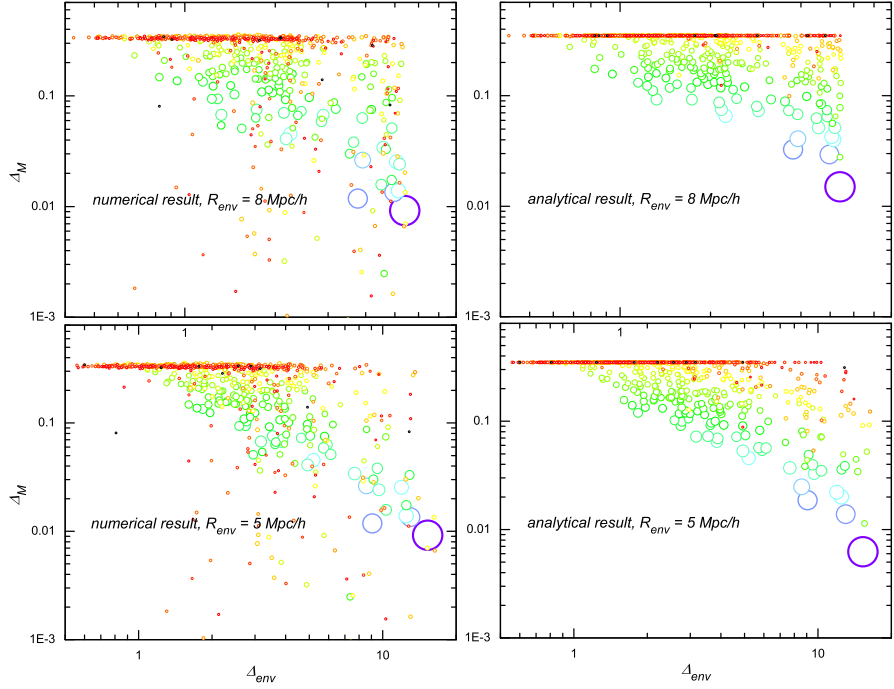}
\caption{(Colour Online) Screening of dark matter halos as a function of the environment and halo mass for the model with $|f_{R0}|=10^{-6}$. The horizontal axis of is $\Delta_{\rm env}$, the matter overdensity of a sphere (the environment) centred at each halo, with comoving radius $R_{\rm env}$ as indicated in the panel. The vertical axis is $\Delta_{M}$, the ratio between the magnitudes of the fifth force and gravity at the surface of a halo. Each circle represents a halo, and the halo's mass is illustrated by both the size (increasing size for increasing mass) and colour (from red to violet for increasing mass) of the circle. The left (right) panels are numerical (analytical) results (see text for a detailed description).}
\label{fig:haloenv}
\end{figure*}

In the $f(R)$ gravity theory, the thin-shell expression Eq.~(\ref{eq:thin_shell}) can be translated into the following equation,
\begin{eqnarray}\label{eq:dror}
\frac{\Delta R}{R} \approx \frac{f_{R,\rm in}-f_{R,\rm out}}{\gamma^2\kappa\rho_{\rm in}R^2},
\end{eqnarray}
by using the relationship between $\sqrt{\kappa}\varphi$ and $f_R$, and the fact that $\sqrt{\kappa}\varphi\sim|f_R|\ll1$. The ratio between the magnitudes of the fifth force and gravity can be approximately estimated as \citep{le2011}
\begin{eqnarray}\label{eq:Delta_M}
\Delta_M &=& \frac{\gamma^2}{2}\times\min{\left\{\frac{3\Delta R}{R},1\right\}}\nonumber\\
&=& \frac{1}{3}\times\min{\left\{\frac{3\Delta R}{R},1\right\}}.
\end{eqnarray}
$\Delta_M$ has a maximum value of $1/3$ as expected, and it can be analytically estimated by calculating $\Delta R/R$ from Eq.~(\ref{eq:dror}) as follows:
\begin{enumerate}
\item Given a halo's mass and virial radius we can compute the average $\rho_{\rm in}$ and therefore $f_{R,{\rm in}}$;
\item $\rho_{\rm out}$, the environmental density, can be estimated by computing the average density of a sphere with a radius $R_{\rm env}$ centring on the concerned halo \citep{le2011}. Then $f_{R,{\rm out}}$ follows straightforwardly.
\end{enumerate}
For the $f(R)$ model considered here, we have
\begin{eqnarray}
f_{R,{\rm in}} = \frac{\left(1+4\frac{\Omega_\Lambda}{\Omega_m}\right)^2}{\left(\tilde{\rho}_{\rm in}+4\frac{\Omega_\Lambda}{\Omega_m}\right)^2}f_{R0},\\
f_{R,{\rm out}} = \frac{\left(1+4\frac{\Omega_\Lambda}{\Omega_m}\right)^2}{\left(\tilde{\rho}_{\rm out}+4\frac{\Omega_\Lambda}{\Omega_m}\right)^2}f_{R0},
\end{eqnarray}
at $z=0$ where $\tilde{\rho}_{\rm out(in)}\equiv\rho_{\rm out(in)}/\bar{\rho}$. So we get $\Delta_M$ from Eq.~(\ref{eq:Delta_M}), with
\begin{eqnarray}\label{eq:Delta_M_ana}
\frac{\Delta R}{R} &\approx&  \frac{\left(1+4\frac{\Omega_\Lambda}{\Omega_m}\right)^2}{2 \tilde{\rho}_{\rm in}\Omega_m\left(RH_0\right)^2}|f_{R0}|\nonumber\\
&\times& \left[\frac{1}{\left(\tilde{\rho}_{\rm out}+4\frac{\Omega_\Lambda}{\Omega_m}\right)^2}-\frac{1}{\left(\tilde{\rho}_{\rm in}+4\frac{\Omega_\Lambda}{\Omega_m}\right)^2}\right].
\end{eqnarray}
This could be measured for each halo from the $N$-body simulations, as is shown in Fig.~\ref{fig:haloenv}. Note that the thin-shell condition for dark matter halos have been studied using $N$-body simulations by \citep{s2010}, but here for the first time we have checked the environmental effects and compared with analytical results.

The left panels of Fig.~\ref{fig:haloenv} show the dependence of $\Delta_M$ on the halo mass (illustrated by the size and colour of the symbols) and the environment matter density (horizontal axis). The results are measured from our $N$-body simulations using two values of $R_{\rm env}$, respectively $8h^{-1}$Mpc (upper left) and $5h^{-1}$Mpc (lower left). There are several interesting features:
\begin{enumerate}
\item Massive halos mostly reside in overdense regions, as could be seen from the correlation between the size of the symbols and the horizontal axis. This is as expected, because only in those regions there are enough particles to form large halos;
\item With the same environmental density, small halos are less screened (have bigger $\Delta_M$), which agrees with the analysis above;
\item For halos with comparable mass, those in overdense regions are more strongly screened, because of the environmental effect;
\item Only very few halos reside in underdense regions, and those are mostly small halos. This is because most particles in those regions have been pulled away. Note that the halos in those regions are essentially unscreened;
\item The dependence of the results on $R_{\rm env}$ is fairly weak, indicating that the exact value of $R_{\rm env}$ in the definition of the environment is not crucial.
\end{enumerate}

In the right panels of Fig.~\ref{fig:haloenv} we have shown the analytical approximation obtained by using Eq.~(\ref{eq:Delta_M_ana}). We can find that the the analytical result agrees with the $N$-body simulation result quite well, and once again there is no sensitive dependence on $R_{\rm env}$. This means that the analytical approximation Eq.~(\ref{eq:Delta_M_ana}) can well describe the nonlinear behaviour of the fifth force in $f(R)$ gravity.

\begin{figure*}
\includegraphics[scale=0.45]{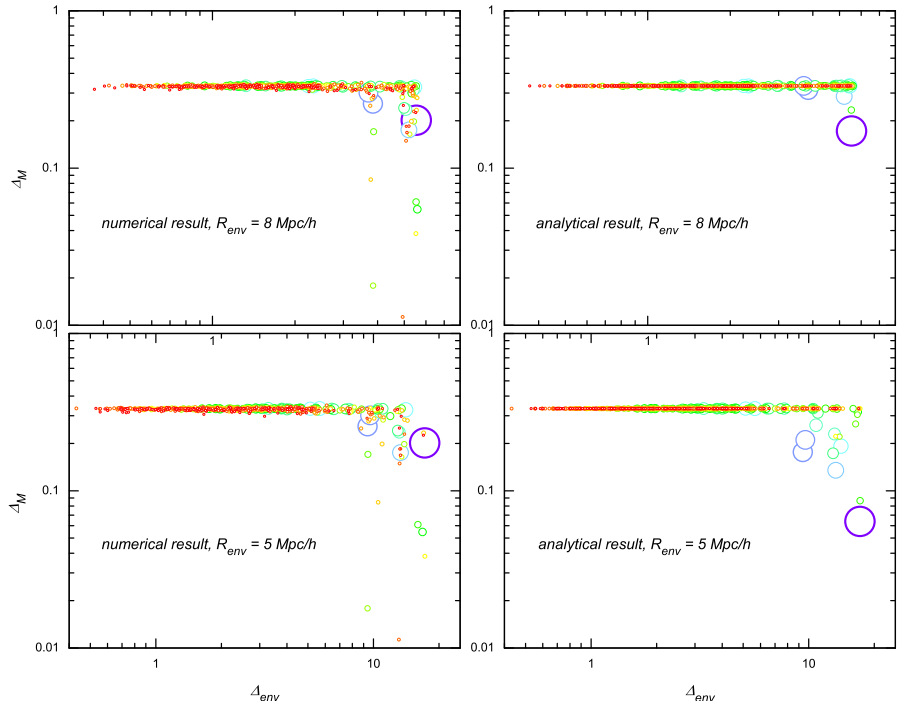}
\caption{(Colour Online) The same as Fig.~\ref{fig:haloenv}, but for the model with $|f_{R0}|=10^{-5}$.}
\label{fig:haloenv2}
\end{figure*}

In Fig.~\ref{fig:haloenv2} we show the  same results for a different $f(R)$ model, with $|f_{R0}|=10^{-5}$. Again we could see that the analytical and numerical results agree well. Note that here all halos but the very massive ones are unscreened. Figs.~\ref{fig:haloenv}, \ref{fig:haloenv2} lend supports to the simple excursion-set model of \citet{le2011} for studying structure formation in the chameleon-type scalar field models.

Of course, the agreement between the analytical and simulation results is not perfect, although it is fairly good statistically. For $|f_{R0}|=10^{-5}$, the analytic thin shell condition underestimates the fifth force especially with $R_{\rm env}=5h^{-1}$Mpc. This is probably because the Chameleon mechanism becomes weaker for larger $|f_{R0}|$ and the environmental effect becomes more non-local. Thus we need a larger $R_{\rm env}$ to capture the environmental effect correctly.

Another possible reason for the slight discrepancy between analytical and numerical results is that the dark matter halos are not rigorously spherical, which we have assumed when applying the thin-shell condition; we have tried to search for the possible correlation between the ellipticity of the halos and $\Delta_M$, but we didn't find any evidence, which means that using the spherical thin shell condition is indeed a good approximation.

\section{Voids}

\label{sect:voids}

\begin{figure*}
\includegraphics[scale=0.5]{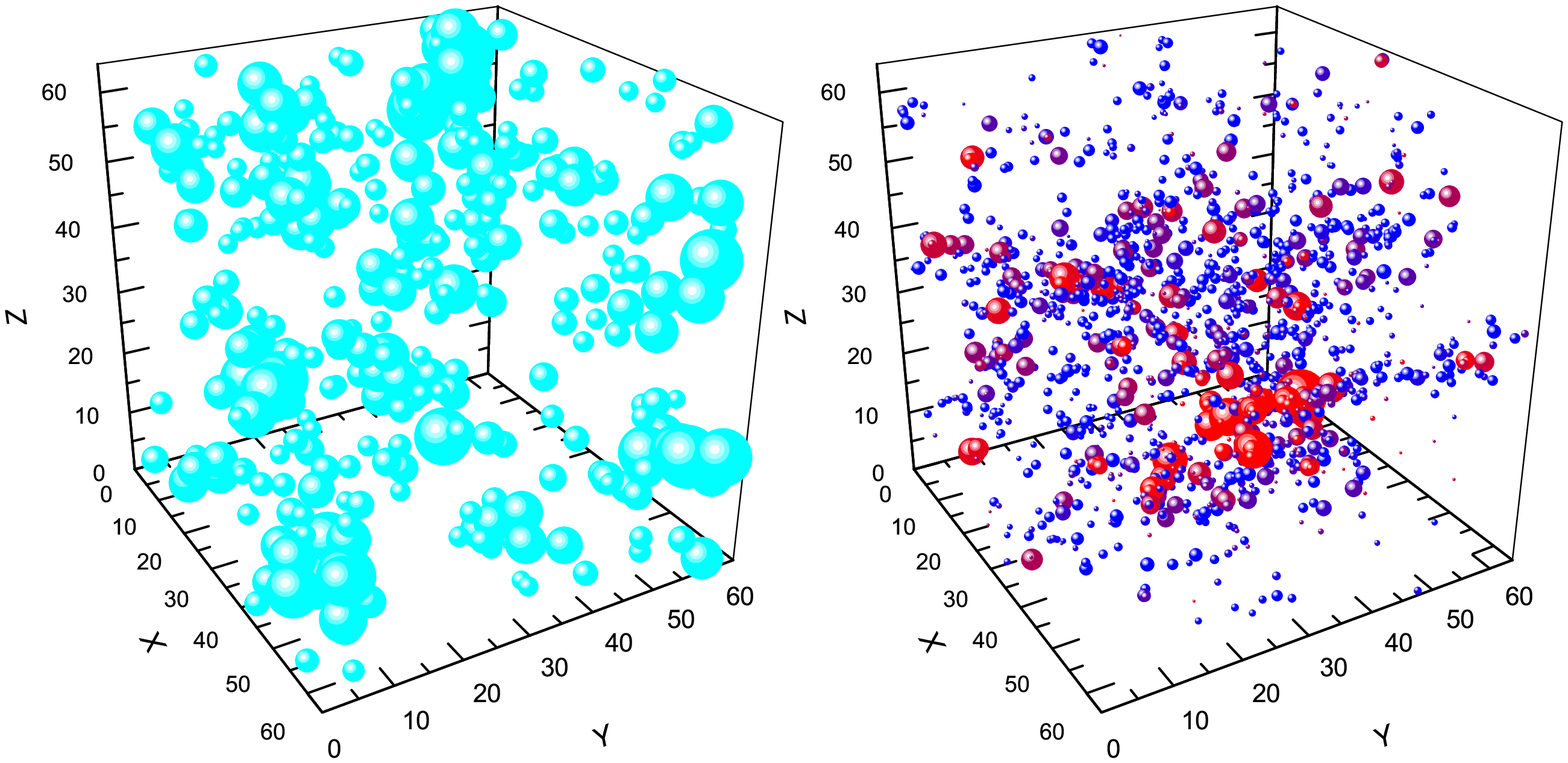}
\caption{(Colour Online) A visualisation of the distributions of voids and dark matter halos in one of our $|f_{R0}|=10^{-6}$ simulations. Each bubble represents a protovoid (left panel, where the bubble size characterises the size of the protovoid) and halo (right panel, where the bubble size characterises the halo's mass). The colour of a halo represents the screening of that halo, from strongly screened (red) to unscreened (blue).}
\label{fig:visualisation}
\end{figure*}

Next let us turn to the underdense regions, or voids, in the $f(R)$ gravity.

The voids are identified using {\tt VAMSUR} (Voids As Merged Spherical Underdense Regions) code developed by \citet{li2011}. The basic idea is to first find the low-density spherical regions (protovoids) and then merge them to form irregularly-shaped voids using a given algorithm. In this work we have chosen $\delta<-0.8$ as the definition of voids.

\begin{figure}
\includegraphics[scale=0.35]{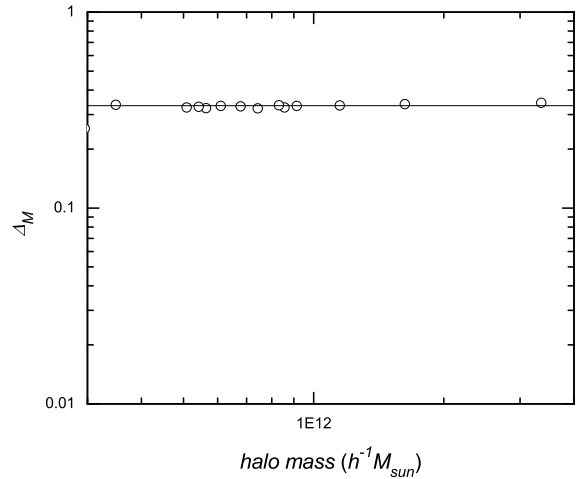}
\caption{Screening of dark matter halos (the open circles) inside voids in one of our $|f_{R0}|=10^{-6}$ simulations. Only halos more massive than $2\cdot10^{12}M_{\odot}$ are shown for resolution considerations. The solid line is $\Delta_M=1/3$ and we can see that these halos are completely unscreened.}
\label{fig:void_halo_screening}
\end{figure}

In Fig.~\ref{fig:visualisation} we show the protovoids (left panel) and dark matter halos (right panel) identified in one of our simulation boxes for the model with $|f_{R0}|=10^{-6}$. We can see that:
\begin{enumerate}
\item Most dark matter halos (in particular the more massive ones) distribute in regions where few protovoids can be identified, and vice versa, which is a trivial test of the code and numerical results.
\item Near the voids the dark matter halos are less screened (denoted in blue) while far from the voids they can be well screened (in red).
\item Halos inside voids are all unscreened as is also shown in Fig.~\ref{fig:void_halo_screening}.
\end{enumerate}
These observations agree with our expectation very well.

\begin{figure}
\includegraphics[scale=0.35]{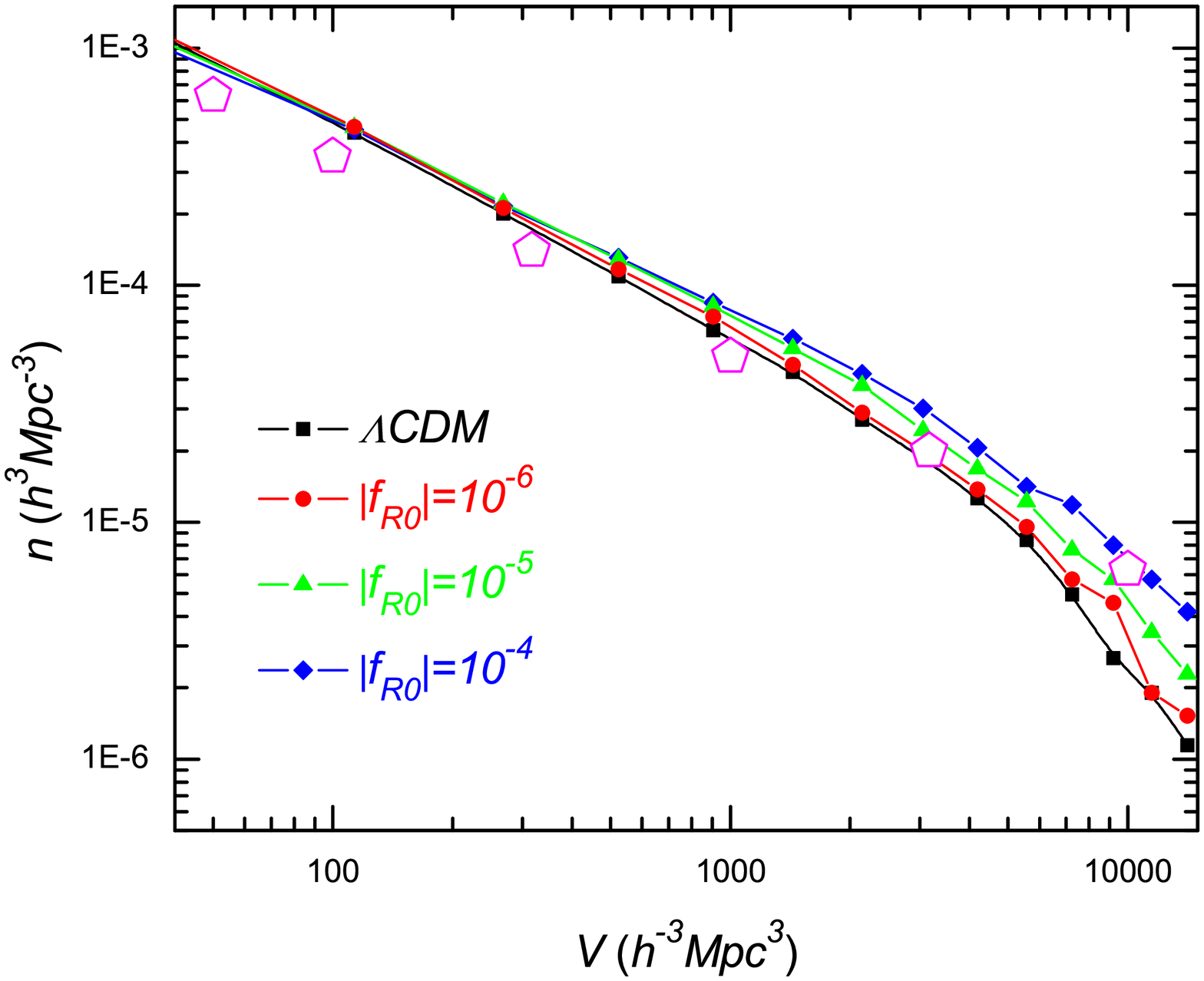}
\caption{(Colour Online) The void number density as a function of a volume. The black squares, red circles, green triangles and blue diamonds are from the $\Lambda$CDM simulation and $f(R)$ simulations with $|f_{R0}|=10^{-6}$, $10^{-5}$, $10^{-4}$ respectively. Each curve is the averaged result of ten realisations. The magenta pentagons are results for $\Lambda$CDM from \citet{csdgy2005} for consistency check. All results are at $a=1$.}
\label{fig:void_nd}
\end{figure}

As mentioned above, in the $f(R)$ gravity the fifth force helps evacuate the low density regions, which results in more large voids than in $\Lambda$CDM. To see this more clearly, one could plot the number density of voids as a function of their effective volumes, and this is shown in Fig.~\ref{fig:void_nd} (see figure caption for details). Here we can see the clear trend: increasing $|f_{R0}|$, which makes the fifth force less suppressed from earlier times, produces more large voids. For example, a $f(R)$ universe with $|f_{R0}|=10^{-5}$ ($10^{-4}$) has twice (four times) as many voids with effective radius $15h^{-1}$Mpc as a $\Lambda$CDM universe does, and the relative difference in the abundance for larger voids is even larger.

Very large voids and very massive dark matter halos are rare objects in a $\Lambda$CDM Universe, and we have seen that both of them are more abundant in the $f(R)$ universes. 
This is the reason why cluster abundance gives the strongest constraints on $|f_{R0}|$ with current observations \citep{svh2009,fsh2011,lssh2010}.
Compared with the halos, the modified gravity effect is more pronounced on the voids. This is because in $f(R)$, gravity is maximumly enhanced in voids due to the presence of the fifth force. At $z=0$, the model with $|f_{R0}|=10^{-5}$ predicts $\sim$30\% more halos with mass $\sim5\times10^{14}M_{\odot}$ than $\Lambda$CDM \citep{zlk2011}, while it predicts twice as many voids of size $\sim15000h^{-3}$Mpc$^3$.

Due to the limitation of our simulation box size, we do not have voids with radius larger than $\sim15h^{-1}$Mpc. However, there is an abundance of such large voids observed. For example, using the SDSS DR7, \citet{pvhcp2011} identified about 1000 voids in the northern galactic hemisphere with radii $>10h^{-1}$Mpc; the largest and median radii in their void catalogue are $30h^{-1}$Mpc and $17h^{-1}$Mpc, respectively. Those voids have an edge density contrast of $\delta<-0.85$. They find that their observations agree quite well with the $\Lambda$CDM simulations, which means that their data could place strong constraints on the $f(R)$ gravity, making voids a promising tool to study the physics of the accelerated cosmic expansion and large-scale structure formation. Of course, their voids are identified by looking at galaxies in the survey, and to make direct comparison with their data we have to generate galaxy catalogues in the $f(R)$ gravity. We will leave this to future work.

\section{Summary and Conclusions}

\label{sect:con}

In this paper, we studied the over- and underdense regions, namely, the distribution of halos and voids, in $f(R)$ gravity simulations. By comparing the probability distribution function of the density contrast $\delta$ for $f(R)$ models with that for GR, we find that there are far more voids in $f(R)$ gravity than that in GR. For example, the numbers of voids with an effective radius of 15$h^{-1}$Mpc are twice and four times as many as those in GR for $f(R)$ models with $|f_{R0}|=10^{-5}$ and $10^{-4}$ respectively. This in principle provides a new means to test GR observationally using the upcoming data. We also find that halos near the voids are less screened and experience stronger gravity. Especially, halos inside the voids are all unscreened in our simulations. This confirms the expectations that small galaxies inside voids provide us the best place for testing modification of gravity.

On the other hand, the overdense regions, ie, the distribution of dark matter halos, can provide important information for the GR test as well. In this work, we utilised the thin-shell condition developed in \citep{kw2004} and \citep{le2011}, and analytically predicted the fractional difference between the lensing mass and the dynamical mass of dark matter halos, $\Delta_M$, as a function of the environment. As we found, the analytic result agrees very well with the simulation result, which means that the thin-shell condition is a good approximation for $f(R)$ gravity. This has important applications for the semi-analytic halo model building for $f(R)$ gravity, which is crucial for realistic constraints of $f(R)$ models using observations.

\section*{Acknowledgments}
BL is supported by the Royal Astronomical Society, Queens' College and Department of Applied Mathematics and Theoretical Physics of University of Cambridge, and thanks the Institute of Cosmology and Gravitation of the University of Portsmouth for its host when part of this work was done. GBZ and KK are supported by STFC grant ST/H002774/1. KK acknowledges supports from the European Research Council and the Leverhulme trust.

\label{lastpage}

\end{document}